\documentstyle[prl,multicol,aps,epsf]{revtex}
\begin{document}

\title{Origin of Multikinks in Dispersive Nonlinear Systems}

\author{Alan Champneys$^1$ and Yuri S. Kivshar$^2$}

\address{$^1$Department of Engineering Mathematics, University of Bristol,
Bristol BS8 1TR, UK \\ $^2$Optical Sciences Centre, Australian National
University, Canberra ACT 0200, Australia}

\maketitle

\begin{abstract}
We develop {\em the first analytical theory of multikinks} for strongly
{\em dispersive nonlinear systems}, considering the examples of the weakly
discrete sine-Gordon model and the generalized Frenkel-Kontorova model with a
piecewise parabolic potential.  We reveal that there are
no $2\pi$-kinks for this model, but there exist {\em discrete sets} 
of $2\pi N$-kinks for all $N>1$. We also show their bifurcation 
structure in driven damped systems.
\end{abstract}


\begin{multicols}{2}
\narrowtext

Non-equilibrium dynamics of many physical systems can be characterized by
the creation and motion of topological excitations or defects.
In particular, when a nonlinear system possesses a degeneracy of its ground
state, such excitations are {\em kinks}, the simplest and probably most
studied nonlinear modes.  The concept of kinks is vital for many physical
problems such as dislocation and mass transport in solids, charge-density
waves, commensurable-incommensurable phase transitions, conductivity,
tribology, Josephson transmission lines, etc  \cite{braun_review}.

In application to problems in solid state physics, the kink's motion
is strongly affected by the inherent lattice discreteness. Earlier
numerical simulations \cite{kruskal} of the kink's motion in a lattice
described by the discrete sine-Gordon (SG) equation, also known as the
Frenkel-Kontorova (FK) model \cite{braun_review}, demonstrated a
number of interesting features not observed in the dynamics of
solitons of integrable (both continuous and discrete) models. In
particular, Peyrard and Kruskal \cite{kruskal} found that a single
kink becomes unstable when it moves in a discrete lattice at
sufficiently large velocity, whereas two (or more) kinks are stable
and propagate as {\em multikinks}.  The former effect is associated
with resonant interaction between a kink and radiation \cite{ust}, and
resonances are even observed experimentally \cite{zant}. In contrast,
the latter phenomenon, i.e. the formation of multikinks, {\em
``... has no clear analytical explanation yet''} (see
\cite{braun_review}, p. 25).

Recently, different physical systems have been studied {\em numerically} where
multikinks are found to play an important role. For example, multikinks are
responsible for {\em a mobility hysteresis} in a damped driven
commensurable chain of atoms \cite{braun}. In arrays of Josephson
junctions, instabilities of fast kinks lead to the generation of {\em
bunched fluxon states} also described by multikink modes \cite{ustinov}.

The main purpose of this paper is to provide the first step towards {\em
an analytical theory of multikinks} in strongly dispersive nonlinear
nonintegrable systems, including the analysis of the existence
and codimension of $N-$kink states.  In particular, we consider a weakly
discrete SG model and demonstrate the existence of {\em a finite number of
multikinks} due to a higher-order dispersion. We also find {\em analytical
solutions for multikinks} and describe the effect of an external field and
damping on their existence and qualitative features.

We consider the dynamics of a commensurable chain of atoms in a periodic
substrate potential. In a normalized form, the equations of motion for the
atomic displacements $u_n$ can be written as
\[
\ddot{u}_n = V_{\rm int}^{\prime} (a_0 + u_{n+1}-u_n) - V^{\prime}_{\rm
int} (a_0 + u_n - u_{n-1}) - W^{\prime}_{\rm sub}(u_n),
\]
where $V_{\rm int}(u)$ is an effective interaction potential with the
equilibrium distance $a_0$, and $W_{\rm sub}(u)$ is a substrate potential
with period $a$.  For small anharmonicity, i.e. when $|u_{n+1} -u_n| \ll
a_0$,  the potential $V_{\rm int}(u)$ can be expanded into a Taylor series
to yield (see details in Ref. \cite{braun_review}):
$\ddot{u}_n - g (u_{n+1} + u_{n-1} - 2u_n) + W_{\rm sub}^{\prime}(u_n) = 0$,
 where $g \equiv V_{\rm int}^{\prime \prime}(a_0)$.  In the quasi-continuum
limit, taking into account a higher-order dispersion, we obtain the
normalized equation
\begin{equation}
\label{eq2}
u_{tt} - u_{xx} - \beta u_{xxxx} + W_{\rm sub}^{\prime}(u) = 0,
\end{equation}
where $W(u)$ has rescaled period $2\pi$ and,
for harmonic interaction, $\beta = a^2/12$.

Equation (\ref{eq2}) takes into account the effect of lattice discreteness
through a fourth-order dispersion term, and for $\beta=0$ and
$ W_{\rm sub}^{\prime}(u) = \sin \, u$, it transforms into the well-known
exactly integrable SG equation that has an analytical solution for {\em a
single $2\pi$-kink} moving with velocity $v$, $u = 4 \tan^{-1} \{
\exp [(x-vt)/\sqrt{1-v^2}] \}$. Similar kinks exist for a rather general
topology of the substrate potential $W_{\rm sub}(u)$ \cite{braun_review}.
However, our aim in this paper is to study {\em a new class of localised
solutions} of Eq. (\ref{eq2}) for $\beta \neq 0$ in the form
of $2\pi N$-multikinks for $N>1$.

First, following the original study of Peyrard and Kruskal \cite{kruskal},
we consider the harmonic substrate potential
\begin{equation}
\label{sine}
W_{\rm sub}(u) = 1 - \cos \, u \,.
\end{equation}
We look for kink-type localised solutions of Eq. (\ref{eq2}) that
move with velocity $v$ $(v^2 <1)$, i.e. we assume $u(z) = u(x-vt)$.
Linearizing Eq. (\ref{eq2}) and taking $u(z) \sim e^{\lambda z}$, we find
eigenvalues $\lambda$ of the form,
\[
\lambda^2 = \frac{1}{2\beta} [(v^2 - 1) \pm \sqrt{(1-v^2)^2 + 4\beta}],
\]
so that for $\beta >0$ there  always exist {\em two real} and {\em two purely
imaginary} eigenvalues.  Thus,  the origin $u=0$ is a saddle-centre point
and hence kinks, which are homoclinic solutions to $u=0$ (mod $2\pi$),
should occur for {\em isolated values} of $v$ for fixed $\beta$
(see Refs. \cite{MiHoOR:98,Ch:98}). That is
they are of codimension one.  Moreover, this codimension is only true
if the solutions are themselves reversible, that is invariant under
one of the transformations:
$$
\begin{array}{l} {\ R_1: \quad  u (\mbox{mod} \: 2 \pi) \to -u  (\mbox{mod}
\: 2 \pi), \; u^{\prime \prime} \to - u^{\prime \prime}, \quad t \to
-t,}\\*[9pt] {\displaystyle
R_2: \quad   u^{\prime} \to - u^{\prime}, \quad u^{\prime \prime \prime}
\to - u^{\prime \prime \prime}, \quad t \to -t,}
\end{array}
$$
where prime stands for differentiation with respect to $z$.

\begin{figure}
\epsfxsize 6.5cm
\centerline{\epsffile{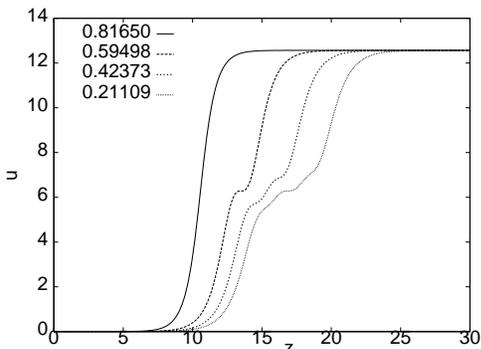}}
\caption{The four $4\pi$-kink solutions of Eq. (\ref{eq2}) with $W_{\rm
sub}^{\prime}(u) = \sin \, u$  propagating at the given velocities.}
\end{figure}

To find {\em all solutions of this type}, first we fix $\beta=1/12$, which
corresponds to $a=1$. Then, we perform numerical shooting on the
ordinary differential equation for $u(z)$ using a
well-established Newton-type
method for homoclinic/heteroclinic trajectories in reversible systems
(see \cite{CS}). The first result is that there exists {\em 
no $2\pi$-kink solution at all}, except in the artificial limit $v^2 \to 
-\infty$ (see comment below). Instead, 
we find a discrete family of $4\pi$-kinks;
specifically there exist {\em only four such solutions} at four
different values of $v$. The first solution has an analytical form
\cite{BoKo:94}
\begin{equation}
\label{exact}
u(z)= 8 \tan^{-1} \exp \left \{ (3\beta)^{1/4}z \right\},
\end{equation}
where $v^2 = 1- 2\sqrt{\beta/3}$, i.e. for our choice of $\beta$,
$v_{4\pi}^{(1)}=\sqrt{2/3}$. Other values are:
$v_{4\pi}^{(2)}=0.59498\ldots$,  $v_{4\pi}^{(3)}=0.42373\ldots$, and
$v_{4\pi}^{(4)}=0.21109\ldots$.  All these solutions are presented in Fig.
1. We may regard this discrete family as part of an infinite sequence of
bound-states of two $2\pi$-kinks that converges to the limit of
infinite separation at a value of $v^2<0$.
Actually the key parameter is $\mu= 1- v^2$, and further numerical
evidence reveals that the bound states converge to $\mu= \infty$ at which
value a $2\pi$-kink exists only formally.

In addition to the $4\pi$-kinks, numerics further reveals 
$v$-values at which $2\pi N$-kinks occur for all $N>2$. 
Figure 2 shows several examples of
$6\pi$- and $8\pi$-kinks.  According to  a dynamical systems theory result
\cite{MiHoOR:98}, on the existence of bound states of homoclinic solutions to
saddle-centre equilibria in reversible Hamiltonian systems, again thinking of
the $4\pi$-kinks as bound states of $2 \pi$-kinks, one should expect to see
{\em precisely two $6 \pi$-kinks for each $4 \pi$-kink}. These would occur
at $v_{6\pi}^{(i),\pm}$ satisfying
$v_{6\pi}^{(i)-} < v_{4\pi}^{(i)}< v_{6\pi}^{(i)+}$; all eight of which
are depicted in Fig.\ 2(a).
Moreover,
there would be two {\em infinite sequences of $8\pi$-kinks} at
$v_{8\pi}^{(i,j)\pm}$ such that
$v_{8\pi}^{(i,j)-} \to v_{4\pi}^{(i)}$ from below as
$j \to \infty$ and $v_{8\pi}^{(i,j)+} \to v_{4\pi}^{(i)}$ from above.
{\em Our numerical simulations have revealed precisely this structure of
all multikink families}.

\begin{figure}
\epsfxsize 7.0cm
\centerline{\epsffile{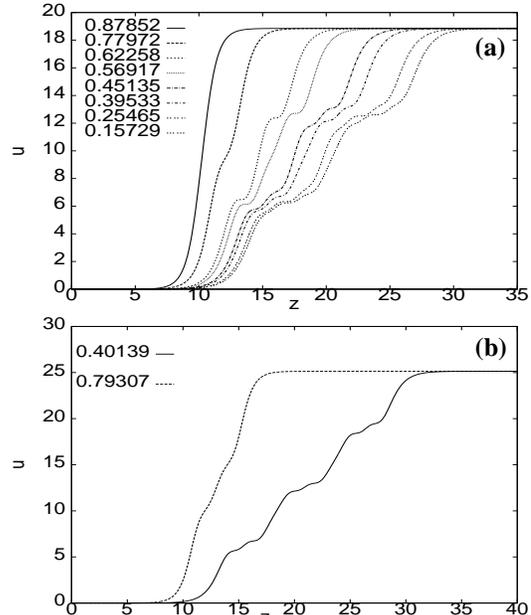}}
\caption{Examples of (a) $6 \pi$- and  (b) $8 \pi$-kinks
of Eq. (\ref{eq2}) with the potential (\ref{sine}) at given velocities.}
\end{figure}

\begin{figure}
\epsfxsize 6.5cm
\centerline{\epsffile{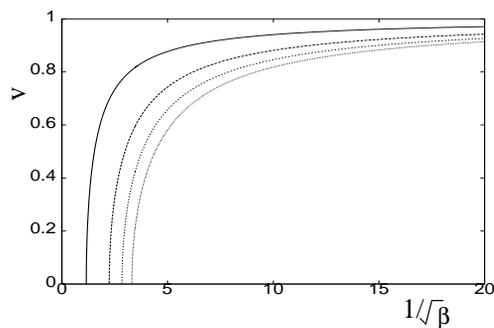}}
\caption{Two-parameter continuation of $4\pi$-kink solutions for the model
(\ref{eq2}) with the harmonic nonlinearity (\ref{sine}).}
\end{figure}

Finally, it appears that the above structure is {\em largely independent}
of $\beta$. Figure 3  shows the results of continuation
(using the method for homo/heteroclinic orbits in the
boundary-value software AUTO \cite{AUTO})
of the four $4\pi$-kinks in the $(v, 1/\sqrt{\beta})$ plane. These
curves are {\em almost identical} to those obtained numerically in Ref.
\cite{kruskal}
for the discrete SG equation.  Note that no curve passes through $v=1$,
they only
reach there asymptotically as $\beta \to 0$. In the process the slope of
each kink at its midpoint steepens, so that the solution becomes singular
in the limit.

It is important that the above numerical results may be verified by the
construction of {\em exact solutions in closed form} when the substrate
potential is
approximated by a piecewise parabolic potential that generates in Eq.
(\ref{eq2}) the effective force,  $W_{\rm sub}^{\prime}(u) =$
$$
\label{pw}
\left \{ \begin{array}{l} {\small u - 2n\pi \: : (2n-1)\pi + \pi/2 < u
< 2n\pi + \pi/2},  \\
{\small (2n+1)\pi -u \: : 2n\pi + \pi/2 < u < (2n+1)\pi + \pi/2 }.
\end{array} \right.
$$
For simplicity we fix $\beta=1/12$ and then look for kinks moving
with velocity $v$ ($\mu = 1-v^2$), by solving the
piecewise-linear equation for $u(z)$.
This defines a four-dimensional dynamical
system in the phase space $(u, u^{\prime}, u^{\prime \prime},
u^{\prime \prime \prime}) \in (-\pi/2,3\pi/2] \times {\bf R}^3$. The
phase space is separated into two distinct domains:
\[
\mbox{\underline{Region 1}}: |u| < \pi/2, \quad
\mbox{\underline{Region 2}}:  \pi/2 < u < 3\pi/2.
\]
$4\pi$-kinks can be constructed by first noticing that,
in order to be of codimension-one (i.e.  occur at
isolated $v$-values),  they should be reversible under the
transformation $R_1$ above. Since we
can always translate by multiples of $2\pi$, we
look for solutions which satisfy, for some unknown $z_2$,
the conditions: $ u(-\infty)
\rightarrow 0$, $u(z_2)=2\pi$, and $u^{\prime \prime}(z_2)=0$,
so that $u(z)$ is in Region 1 for all $z<0$, in Region 2 for
$0 < z < z_1$, for some unknown $z_1<z_2$, and is in Region 1 again for
all $z_1<z<z_2$.

The boundary condition can be satisfied by noticing that such
solutions at $z=0$ (the first point of transition between Regions 1
and 2) satisfy $u(0)=\pi/2$, $u^{\prime}(0)=\lambda\pi/2$, $u^{\prime
\prime}(0)=\lambda^2\pi/2$, and $u^{\prime \prime
\prime}(0)=\lambda^3\pi/2$, where $\lambda^2 = 6(\sqrt{\mu^2
+2/3\pi}-\mu)$ is the unique real positive eigenvalue of the linear
system in Region 1.  Hence the asymptotic boundary condition at
$z=-\infty$ in Region 1 becomes an initial condition at $z=0$ for $u$
in Region 2.  The general solutions in Regions 1 and 2 are:
\[
u_1(z)=A_1 e^{\lambda z} + B_1 e^{\lambda z} + C_1 \cos (\omega z) +
D_1 \sin (\omega z)
\]
and, providing $\mu>\mu_{\rm min}:=\sqrt{2/3\pi}$,
\[
u_2(z) = A_2 \cos (\omega_1 z +B_2) + C_2 \cos (\omega_2 z +D_2)
\]
where $\omega^2 = 6(\sqrt{\mu^2 + 2/3\pi}+\mu)$ and $\omega_{1,2}^2 = 6(\mu
\pm \sqrt{\mu^2 -2/3\pi})$, $A_j$, $B_j$, $C_j$, and $D_j$ are unknown
coefficients. Therefore, we can explicitly solve for the coefficients
to find $u_2(z)$ in closed form. This expression defines an
implicit equation for $z_1$; $u_2(z_1)=3\pi/2$. The value of $u_2(z_1)$
and its derivatives then defines initial conditions at $z=z_1$, hence
determining the constants $A_1$, $B_1$, $C_1$, and $D_1$. This in turn
defines $z_2$ implicitly as $u_1(z_2)=2\pi$.  To have a $4\pi$-kink we
additionally require $u_1^{\prime \prime}(z_2)=0$, and so should only
expect to find zeros of this final quantity by varying $v$. Hence we
can define a `test function' for $4\pi$-kinks $K(v;z_1,z_2):=
u_2^{\prime \prime}(z_2)$. Using the above construction, this $K$ can
be written in closed form in terms of $v$, $z_1$, and $z_2$. The
unknown transition points $z_{1,2}$ are the solution to given transcendental
equations, in each case only the first solution of which has meaning.

\begin{figure}
\epsfxsize 8.0cm
\centerline{\epsffile{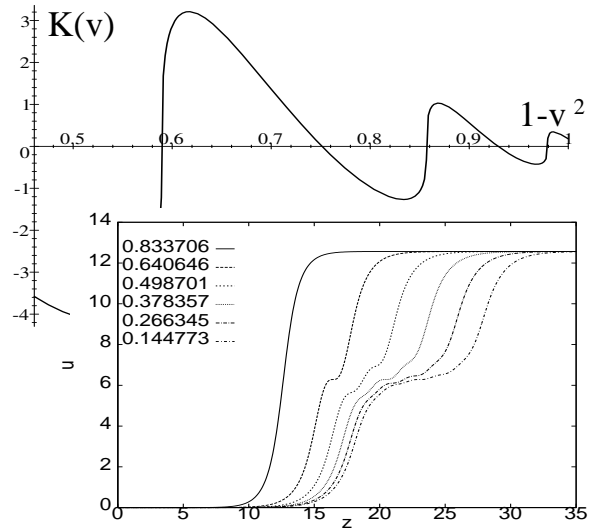}}
\caption{The function $K(v)$ for multikinks of Eq. (\ref{eq2}) with piecewise
parabolic potential. Inset: the corresponding $4\pi$-kinks and their
associated $v$-values.}
\end{figure}

Figure 4 shows a graph of $K$ as a function of $\mu=1-v^2 \in
(\mu_{\rm min},1)$, which has been computed using MAPLE with the
implicit equations solved for their smallest positive
solutions. The five zeros of $K$ correspond to $4\pi$-kinks, graphs of
which are shown in the insert to the figure.  These zeros occur for $v
= 0.64064609$, 0.49870155, 0.37835717, 0.26634472, 0.14477294. It is
also possible to construct solutions for $\mu<\mu_{\rm min}$ in a
analogous manner, but with the solution in Region 2 replaced by one
corresponding to complex eigenvalues. This gives the additional
solution for $v=0.833706$.

In this way, we find {\em analytically} a finite set of $v$-values
giving $4\pi$-kinks for the piecewise parabolic potential model, having {\em
qualitatively the same structure} as the solutions found numerically for
the sinusoidal nonlinearity (\ref{sine}).  One could go on to construct $2\pi
N$-kinks for $N>2$, but the calculations presented already serve to
corroborate the earlier numerical results.

To complete the analysis of the kinks, we would like to mention that the
short-wavelength instability of the {\em nonstationary} continuous model
(\ref{eq2}) due to the term $u_{xxxx}$ can be easily removed by introducing
an equivalent higher-order dispersion via a mixed derivative term
\cite{braun_review,rosenau}.

To analyse the robustness of multikinks in realistic physical systems,
we add to the right-hand side of Eq. (\ref{eq2}) the driven damped
term $F-\delta u_t$, where $F$ is an external DC force and $\delta$ is
a damping coefficient (see, e.g., \cite{braun}).  Importantly,
for each of the kinks so far found, it is possible to use numerical
continuation to trace curves that lie on sheets in the parameter space
$(v, \delta, F)$ corresponding to the existence of multikinks. For
example, taking the explicit $4\pi$-kink solution given by
Eq. (\ref{exact}), a curve was computed at $\beta=1/12$ in the
$(F,v\delta)$-plane with fixed $|v|= \sqrt{2/3}$, reaching a maximum with
respect to $\delta$ at $|v|\delta=0.069326$. Taking the fixed value
$|v| \delta=0.05$ from this curve the locus of kinks in the $(v,F)$-plane
can then be traced out, as depicted in Fig. 5.

\begin{figure}
\epsfxsize 8.0cm
\centerline{\epsffile{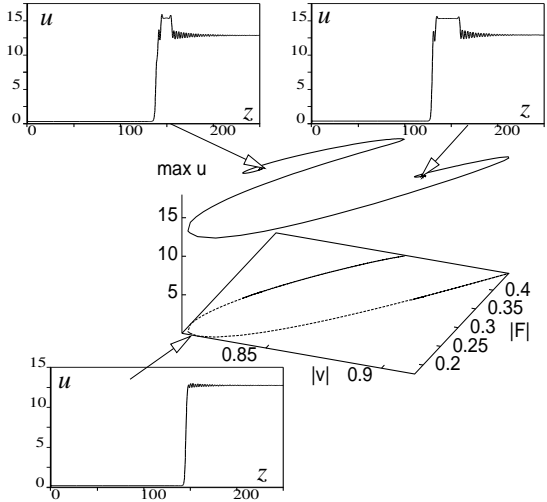}}
\caption{The kink's velocity $v<0$ and maximum amplitude
against external force $F>0$ for the simplest $4\pi$ kink with $v\delta=0.05$.
Note that each limb of the curve spirals back on
itself. The insets illustrate example solutions on the locus.}
\end{figure}

Three interesting features can be noted from this curve. First, all
kinks have developed oscillations around the equilibrium close to
$u=4\pi$. This is because, for $\delta>0$, the corresponding equation
for travelling waves is no longer Hamiltonian or reversible, and the
linearization around the asymptotic value $u_{\star} = \sin^{-1} F$
now has {\em three stable eigenvalues}, two of which have non-zero
imaginary part. These oscillations may be regarded as {\em radiation
that travels at the kink's velocity}, as was earlier observed in
direct numerical simulations \cite{ustinov}. Second, {\em $v$ and $F$ have
opposite sign} for these results.  When $F$ and $v$ have the same sign,
only kinks with {\em non-decaying} oscillations in the tails can be
found.  Third, note that the computed curve ends at a point where a
transition takes place involving a heteroclinic connection with $u
\approx 5\pi$. This suggests that $\pi$-kinks are possible for
sufficiently large $F$.

Finally, we mention that the case $\beta <0$ in Eq. (\ref{eq2}) can also
occur in generalised nonlinear lattices provided we take into account
the next-neighbor interactions, e.g. due to the so-called {\em helicoidal
terms} in nonlinear models of DNA dynamics \cite{helic}. In this
case, the analysis is much simpler and, similar to the nonlocal SG
equations \cite{nonlocal}, leads to the  continuous families of multikinks
parameterized by $v$. From the mathematical point of view, for $\beta <0$
the origin changes from a saddle-center to a saddle-focus, and
rigorous variational principles \cite{Kalies} give
families of stable $2\pi N$-kinks for all $N>1$.

In conclusion, we have developed the first analytical theory of
multikinks in strongly dispersive nonlinear systems, considering the
important examples of the generalized FK model with the sinusoidal and
piecewise parabolic potentials. We have revealed, numerically and
analytically, the existence of discrete sets of $2\pi N$-kinks.  We
believe that general features of multikinks and the physical mechanism for
their formation are similar in many other strongly dispersive nonlinear
models.

Yuri Kivshar thanks O.M. Braun, A.S. Kovalev, B. Malomed, M. Peyrard, and
A. Ustinov for useful discussions. Alan Champneys is indebted to the
Optical Sciences Centre for hospitality and to the UK EPSRC with whom he
holds an advanced fellowship.

\end{multicols}
\end{document}